# Current-driven spin orbit field in $LaAlO_3/SrTiO_3$ heterostructures

Kulothungasagaran Narayanapillai,[1] Kalon Gopinadhan,[1,2] Xuepeng Qiu,[1] Anil Annadi,[2,3] Ariando,[2,3] Thirumalai Venkatesan,[1,2,3,4] and Hyunsoo Yang[1,2,a]

[1]Department of Electrical and Computer Engineering, National University of Singapore, 117576, Singapore

[2]NUSNNI-Nanocore, National University of Singapore, 117411, Singapore

[3]Department of Physics, National University of Singapore, 117542, Singapore

[4]Department of Materials Science and Engineering, National University of Singapore, 117575, Singapore

We demonstrate a current tunable Rashba spin orbit interaction in $LaAlO_3/SrTiO_3$ (LAO/STO) quasi two dimensional electron gas (2DEG) system. Anisotropic magnetoresistance (AMR) measurements are employed to detect and understand the current-induced Rashba field. The effective Rashba field scales with the current and a value of 2.35 T is observed for a dc-current of 200 μA. The results suggest that LAO/STO heterostructures can be considered for spin orbit torque based magnetization switching.

[a] e-mail address: eleyang@nus.edu.sg

Current-induced spin orbit torque (SOT) driven magnetization dynamics have gained much attention due to their applications in logic and memory devices[1-5]. Spin orbit torques which originate from the spin Hall effect or Rashba effect provide efficient ways to control and manipulate the magnetization. In such systems, in-plane current-induced switching[6,7], fast domain wall motion[5,8] and magnetic oscillations[9,10] are reported. In addition to metallic bilayer systems such as ferromagnet (FM)/heavy metal (HM), multilayer systems such as Co/Pd multilayers[11] and magnetically doped topological insulator heterostructures[12] have been investigated. In these systems, the spin orbit torque due to the spin Hall effect originates from heavy metals as the spin orbit coupling strength increases with the fourth power of the atomic number. In contrast, structural inversion asymmetry across the surfaces and interfaces results in strong Rashba torques in two dimensional electron gas (2DEG) systems such as GaAs/GaAlAs,[13] InAs/InGaAs,[14] and $LaAlO_3/SrTiO_3$ (LAO/STO)[15,16]. These 2DEG systems are widely explored for high performance spintronics applications such as spin-transistor, as the gate tunable Rashba field makes such systems an attractive choice.

Recently, there has been a great interest in LAO/STO systems due to the observation of an electric field tunable quasi-2DEG[17,18]. These oxide systems exhibit unusual properties such as the simultaneous existence of magnetism and superconductivity[19], magnetism and Kondo scattering[20], etc. The existence of these competing phenomena is possible only if spin orbit interaction is present. Therefore, understanding the spin orbit interaction at this interface and its tunability is of great importance. The electron gas present in LAO/STO heterostructures is confined within a few nanometers from the interface on the STO side[21]. This structural configuration of the heterostructure breaks inversion symmetry. As a result, the electron gas confined in the vicinity of a polar (LAO)/non-polar (STO) interface experiences a strong electric field directed perpendicular to the conduction plane[22]. The effective electric field is given by the Rashba Hamiltonian $H_R = \alpha_R (\hat{n} \times \vec{k}) \cdot \vec{S}$, where $\vec{S}$ are the Pauli matrices, $\vec{k}$ is the electron wave vector, and $\hat{n}$ is a unit vector perpendicular to the interface. This Hamiltonian describes the coupling of the electrons spin to an internal magnetic field ($\propto \hat{n} \times \vec{k}$) which is experienced in its rest frame. The internal magnetic field is perpendicular to their wave vector and lies in the plane of the interface.

The magnitude of the strong Rashba interaction can be tuned by an external electric field. This presents an advantage to control the Rashba field in this system which is due the special dielectric properties of $SrTiO_3$[15]. Moreover, the observed ferromagnetism[19,23] in LAO/STO systems provides alternative ways to probe spin orbit torques using anisotropic magnetoresistance (AMR)[18,24]. Probing of AMR in a rotating in-plane magnetic field is a powerful tool to detect possible magnetic ordering in LAO/STO systems[24]. The magnetic order formed in this system is found to be dependent on the number of charge carriers, temperature, and film thicknesses[24]. Rashba spin orbit interactions in LAO/STO systems is also probed by AMR[15,25,26], and the Rashba strength has been estimated by applying two dimensional weak (anti)localization theory which is complex and involves many parameters. In this study, we directly utilize AMR measurements with the external magnetic field to quantify the Rasbha field. So far, the Rashba field related studies in this system have been limited to the electric field effect. In this letter, we study the current-induced Rashba field in the LAO/STO system utilizing angular dependent AMR. We find a two-fold MR oscillation for the LAO/STO 2DEG systems. The Rashba field is extracted from the current dependent AMR response and an induced field of 2.35 T is estimated for a dc-current of 200 µA.

The devices are prepared using the following steps. First, device structures with Hall crosses (width of 50 µm) are patterned using a negative tone (MaN 2405) resist on $TiO_2$ terminated STO (001) substrates, which are pre-treated with buffered oxide etch (BOE) and air annealing at 950 °C for 1.5 h, using electron beam lithography (EBL). A blanket deposition of amorphous $AlN_X$ (20 nm) at room temperature by pulsed laser deposition (PLD) is used to cover the STO surface except the patterned area. This step ensures that the surrounding area around the pattern does not conduct, since the STO/AlNx interface does not form any 2DEG. After removing the resist, the sample is annealed at 750 °C for 60 minutes in oxygen and subsequently, a blanket LAO is grown by PLD with an oxygen partial pressure of 1 mTorr at 750 °C. As a result, a 2DEG LAO/STO interface is formed only at the defined device structure. We have utilized 5 unit-cells (uc) of LAO (lattice parameter = 0.39 nm) which implies an LAO thickness of ~ 2 nm. The carrier density of this sample is estimated to be $n_s = 2.5 \times 10^{13}$ cm$^{-2}$ at 4 K. The structure of the film stack is shown in Fig. 1(a). Electrical contact pads were made by wire bonding to the sputtered Ta (4 nm)/Cu (90 nm) contact pads defined by EBL. The Hall crosses are designed parallel to the edges to align them along the crystallographic axis (001).

Measurement schematics utilized to detect AMR and planar Hall effect (PHE) measurements are shown in Fig. 1(b) where lock-in amplifier (LIA) 1 and 2 are utilized to measure AMR and PHE, respectively. Measurements are performed in a physical property measurement system (PPMS) under He atmosphere equipped with a rotating sample probe and a 9 T superconducting magnet. Keithley 6221 is used to source currents, and lock-in amplifiers are utilized to measure the voltage across the channel and the Hall bars.

Figure 1(c) and (d) show the measured AMR and PHE values at 4 K with an applied magnetic field of 9 T. A 20 µA ac-current ($I_{ac}$ = 20 µA, $I_{dc}$ = 0 µA) with a frequency of 13.3 Hz is used for the measurements. Both current (*I*) and magnetic field (*H*) are in the plane of the sample, and the angle ($\theta$) is varied from 0° to 360°. Here, $\theta$ is defined as the angle between the Hall bar direction and the applied field direction as depicted in Fig. 1(b). The AMR data show a clear two-fold oscillation. A phenomenological model is commonly used to describe the AMR and PHE values in typical 3*d* ferromagnetic systems[27-29]. In this model, the resistivity tensor is defined in terms of the direction of current with respect to the applied magnetic field. The resistance is given by the following equation with higher order cosine terms in a cubic symmetric system;

$$R_{XX} = a_0 + a_1 \cos^2(\theta + \phi) + a_2 \cos^4(\theta + \phi) \tag{1}$$

$$R_{XY} = a_3 + a_4 \sin(2\theta + \phi_1) \tag{2}$$

where $R_{XX}$ and $R_{XY}$ are related to AMR and PHE, respectively. The coefficients $a_1$, $a_2$, and $a_4$ arise from the uniaxial and cubic components of magnetization, while $a_0$, $a_3$, $\phi$ and $\phi_1$ are constants introduced to fit the observed signals. The fit of Eq. (1) shows good agreement with the measured AMR data as in Fig. 1(c). Such a two-fold oscillation is a common feature of LAO/STO systems[18,26]. Similarly, the fit of the PHE data with Eq. (2) is shown in Fig. 1(d). The good agreement of the experimental data with the model implies the presence of interacting magnetic moments in the system.

In order to study current-driven Rashba fields, we have studied the current dependence of AMR by varying the dc-current. The external magnetic field ($H_A$), applied during the AMR measurements, is modulated by the current-induced Rashba field ($H_R$) as depicted in Fig. 2(a). Therefore, the resultant magnetic field ($H_{EFF}$) experienced by the channel is given by

$$H_{EFF}^2 = H_A^2 + H_R^2 + 2H_A H_R \cos\theta \tag{3}$$

$$\alpha = \arctan\left[H_R \sin\theta / (H_A + H_R \cos\theta)\right] \tag{4}$$

where $\theta$ ($\alpha$) is the angle between $H_A$ ($H_{EFF}$) and $H_R$ ($H_A$) in Fig. 2(a). AMR in LAO/STO structures as given by Eq. (1) depends on the external magnetic field, $H_A$ in such systems. The coefficients in Eq. (1), $a_1$ and $a_2$, are found to be linearly dependent with the external magnetic field. By incorporating the field dependency in the equation,

$$R_{XX} = b_0 + b_1 H_{EFF} \cos^2(\beta+\varphi) + b_2 H_{EFF} \cos^4(\beta+\varphi) \tag{5}$$

where $b_0$, $b_1$, $b_2$, and $\varphi$ are constants, and $\beta$ is defined as $\beta = \theta - \alpha$. In order to modulate the current induced Rashba field, we increase the dc-current to $I_{dc} = \pm 250$ µA as shown in Fig. 2(b). The measurements are performed with $H_A = 9$ T and 4 K. It is evident that the angular dependency of the AMR with both positive and negative currents is significantly modulated by the introduction of the current-induced Rashba field. The fits of the Eq. (5) to the measured AMR data are shown in Fig. 2(b). The extracted Rashba field values are $H_R = 1.26$ T and -1.48 T for +250 and -250 µA, respectively. By assuming a 2DEG thickness of 7 nm as reported earlier,[30] 250 µA corresponds to a current density of $7.14 \times 10^8$ A/m$^2$. The current-induced Rashba field at the current density $10^{12}$ A/m$^2$ is $1.76 \times 10^3$ T.

The strength of the spin orbit coupling is denoted by $\alpha_R$ in the Rashba Hamiltonian. The correction to the band structure due to atomic spin orbit coupling is expected to be weak because of low atomic number of Ti, therefore, the interface electric field induced Rashba effect is assumed to be dominating the Hamiltonian. Additionally, the atomic spin orbit coupling does not introduce any spin splitting along the momentum axis, however, the Rashba effect does. Therefore, $\alpha_R$ represents the strength of the Rashba spin orbit coupling. The spin relaxation time $\tau_{so}$ is defined as $q_{so}^2 = 1/D\tau_{so}$ in the HLN formalism[15,31]. Furthermore, $\alpha_R$ is related to $q_{so}$ as $\alpha_R = \hbar^2 q_{so}/2m^*$. Effective mass of the electron is taken as $m^* = 3m_0$, where $m_0$ is the mass of a free electron. The spin orbit field, $H_{so}$ is given by $H_{so} = \hbar/4eD\tau_{so}$. By direct substitution, $\alpha_R = \sqrt{\hbar^3 eH_{so}/m^{*2}}$. For $H_{so}$ = 1.48 T, $\alpha$ = 12 meV·Å. The spin splitting energy is given by $\Delta = 2\alpha k_F$, where $k_F$ is the Fermi wave vector. $\Delta$ is estimated to be $\Delta$ = 3 meV, which is comparable to that of reported values[31].

In order to further confirm the current-induced Rashba phenomenon, we have systematically studied the dependency of dc-currents with the electric field. A large low temperature permittivity of STO ($\varepsilon_r > 10^4$) allows a higher modulation of the electric field in a 0.5 mm thick STO substrate through an application of back gate voltages. The resistance of the channel decreases with increasing the bias voltage ($V_g$). In order to probe the real ground state of the system (as there can be traps at the interface which lead to hysteresis under an applied electric field), $V_g$ was swept 5 times from $V_g$ = 50 to -50 V before the start of the measurements. This step reduces the hysteresis and ensures the reproducibility. Furthermore, the amplitude of AMR increases with increasing $V_g$. Therefore, we set $V_g$ = 50 V to get a higher modulation of AMR values in the following subsequent experiments. Figure 3(a) shows AMR measurements for positive dc-current values up to 200 µA with $H_A$ = 8 T. It is clear that with increasing dc-current values, the asymmetry of AMR increases. The fits of Eq. (5) in Fig. 3(a) show good agreement, and correctly capture the features of the curves for every dc current value ($I_{dc}$). Furthermore, we have also performed AMR measurements for $H_A$ = -8 T and the respective fits with Eq. (5) are shown in Fig. 3(b). We can observe that the AMR results in Fig. 3(a) and (b) show a similar behavior with a 180° phase shift. The extracted values of Rashba field ($H_R$) scales linearly with increasing the dc current ($I_{dc}$) as shown in Fig. 4. Both positive and negative applied field ($H_A$) show similar values of $H_R$. The current induced Rashba field decreases to $H_R$ = 0.8 T as the $V_g$ decreases to 10 V for $I_{dc}$ = 200 µA (not shown) which is similar to the reported values[15].

We have further investigated the current-induced Rashba field by sweeping the magnetic field at a fixed angle, $\theta$ = 0°. The AMR with $I_{ac}$ = 20 µA ($I_{dc}$ = 0 µA) as shown in Fig. 5(a) shows symmetry across positive and negative fields. However, when a high dc-current of ±200 µA is applied, we can see an asymmetry in Fig. 5(b). For the positive $I_{dc}$, the current induced Rashba field ($H_R$) aligns along the +y-direction, and it adds to the applied field ($H_A$) when $H_A$ is positive, and opposes if $H_A$ is negative.

The ability to control the Rashba field with electric currents is technologically advantageous which can be directly used in many proposed systems such as magnetic memories and spin orbit torque oscillators. In metallic systems, a maximum value of 1170 Oe is observed for current induced effective fields in the transverse direction for a current density of $10^{12}$ A/m$^2$ in Co/Pd multilayers[11]. However, in LAO/STO systems, the observed Rashba field is in the order of a few thousand Tesla at that current density. Furthermore, the ability to effectively tune the Rashba field with electric fields makes LAO/STO system very attractive.

In summary, we have studied current-induced Rashba fields in LAO/STO systems. Angular dependence of AMR is used to quantify the Rashba field. The current-induced Rashba field modulates with the externally applied magnetic field and its effect is observed in the AMR values. The modulation of the Rashba field with electric fields is also investigated. The demonstration of current-induced Rashba fields in LAO/STO systems, along with the electric field tunability, paves the way for an alternative approach in spin orbit torque related fields.

This research is supported by the National Research Foundation, Prime Minister's Office, Singapore under its Competitive Research Programme (CRP Award No. NRF-CRP12-2013-01).

Figure captions

Figure 1: (a) Schematics of LAO/STO stack layers. (b) Measurement schematics to probe AMR and PHE. (c) AMR data with in-plane rotation ($\theta$) with a fit (line). (d) PHE data with in-plane rotation with a fit (line).

Figure 2: (a) Directions of $H_R$ and $H_A$ with respect to the current ($I$). (b) AMR data with $I_{dc} = \pm 250$ µA and fits.

Figure 3: AMR with various $I_{dc}$ at $H_A = 8$ T (a) and $H_A = -8$ T (b) with fits.

Figure 4: Rashba field ($H_R$) values extracted from the data in Fig. 3 with various $I_{dc}$.

Figure 5: AMR data with $I_{dc} = 0$ (a) and $I_{dc} = \pm 200$ µA (b) at $\theta = 0°$.

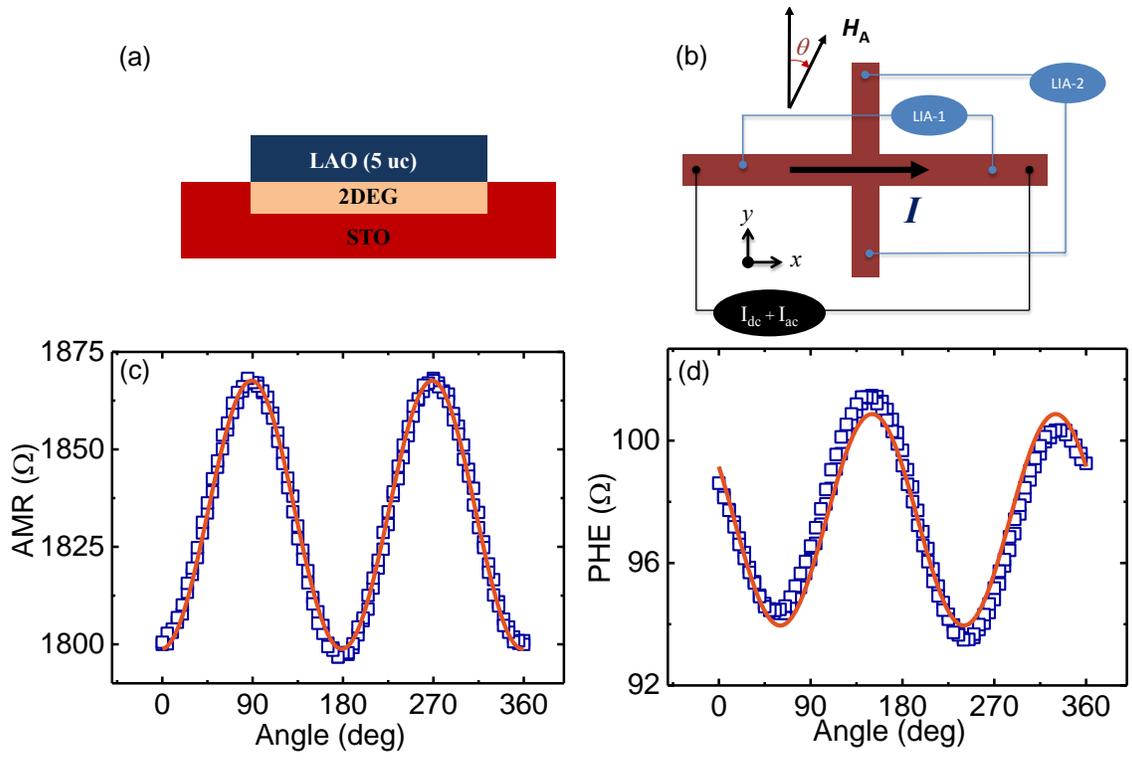

Figure 1

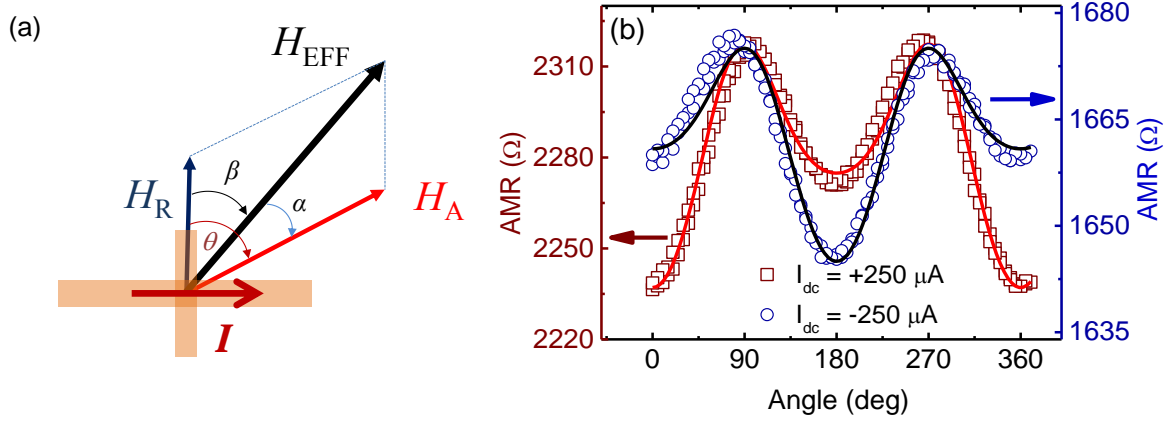

Figure 2

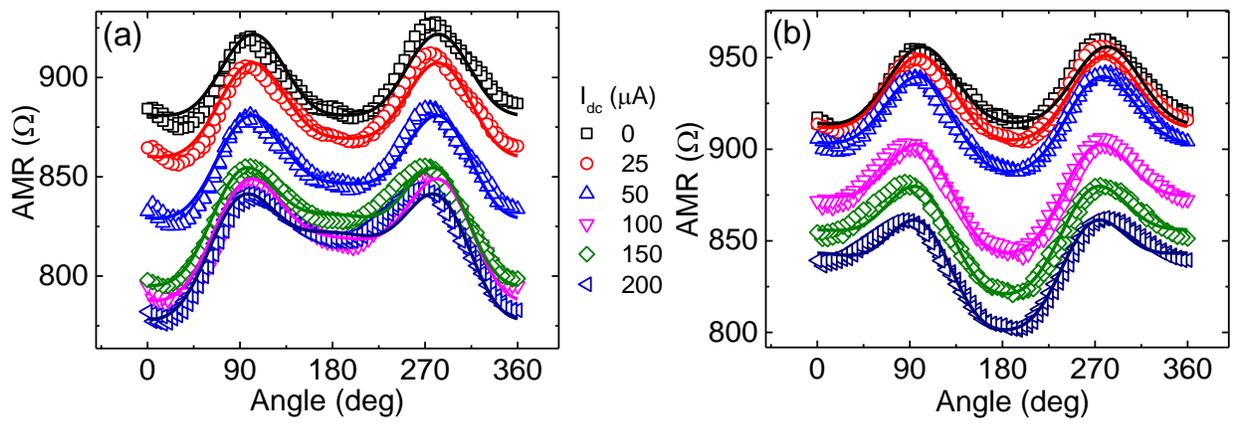

Figure 3

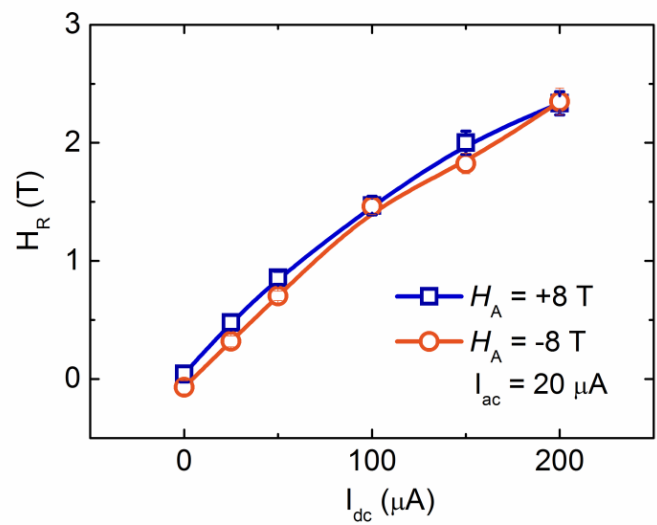

Figure 4

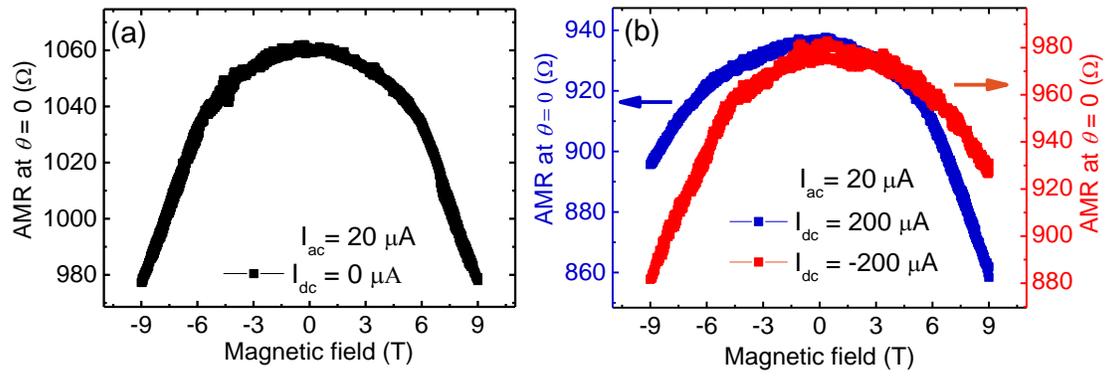

Figure 5